\documentclass[12pt,preprint]{aastex}
\usepackage{natbib}

\begin{document}

\title{Embedded Star Formation in S$^4$G Galaxy Dust Lanes}

\author{
Debra M. Elmegreen\altaffilmark{1},
Bruce G. Elmegreen\altaffilmark{2},
Santiago Erroz-Ferrer\altaffilmark{3,4},
Johan H. Knapen\altaffilmark{3,4},
Yaron Teich\altaffilmark{1},
Mark Popinchalk\altaffilmark{1,5},
E. Athanassoula\altaffilmark{6},
Albert Bosma\altaffilmark{6},
S\'ebastien Comer\'on\altaffilmark{7,15},
Yuri N. Efremov\altaffilmark{8},
Dimitri A. Gadotti\altaffilmark{9},
Armando Gil de Paz\altaffilmark{10},
Joannah L. Hinz\altaffilmark{11},
Luis C. Ho\altaffilmark{12},
Benne Holwerda\altaffilmark{13},
Taehyun Kim\altaffilmark{9,14},
Jarkko Laine\altaffilmark{7},
Eija Laurikainen\altaffilmark{7,15},
Kar\'{i}n Men\'{e}ndez-Delmestre\altaffilmark{16},
Trisha Mizusawa\altaffilmark{17,18},
Juan-Carlos Mu\~noz-Mateos\altaffilmark{17},
Michael W. Regan\altaffilmark{19},
Heikki Salo\altaffilmark{7},
Mark Seibert\altaffilmark{12},
Kartik Sheth\altaffilmark{17}
}
\altaffiltext{1}{Vassar College, Dept. of Physics and Astronomy, Poughkeepsie, NY 12604}
\altaffiltext{2}{IBM Research Division, T.J. Watson Research Center, Yorktown Hts., NY 10598}
\altaffiltext{3}{Instituto de Astrof\'{i}sica de Canarias, V\'{i}a L\'{a}ctea s/n 38205 La Laguna, Spain}
\altaffiltext{4}{Departamento de Astrof\'{i}sica, Universidad de La Laguna, 38206 La Laguna, Spain}
\altaffiltext{5}{Department of Astronomy, Wesleyan University, Middletown, CT 06459}
\altaffiltext{6}{Aix Marseille Universit\'e, CNRS, LAM (Laboratoire d'Astrophysique de Marseille) UMR 7326, 13388, Marseille, France}
\altaffiltext{7}{Department of Physical Sciences/Astronomy Division, University of Oulu, FIN-90014, Finland}
\altaffiltext{8}{Sternberg Astronomical Institute of the Lomonosov Moscow State University, Moscow 119992, Russia}
\altaffiltext{9}{European Southern Observatory, Casilla 19001, Santiago 19, Chile}
\altaffiltext{10}{Departamento de Astrof\'isica, Universidad Complutense de Madrid, Madrid 28040, Spain}
\altaffiltext{11}{MMTO Observatory, The University of Arizona, 933 North Cherry Avenue, Tucson AZ 85721 USA}
\altaffiltext{12}{The Observatories, Carnegie Institution of Washington, 813 Santa Barbara Street, Pasadena, CA 91101}
\altaffiltext{13}{European Space Agency Research Fellow (ESTEC), Keplerlaan 1, 2200 AG Noordwijk, The Netherlands}
\altaffiltext{14}{Astronomy Program, Department of Physics and Astronomy, Seoul National University, Seoul, 151-742 Korea}
\altaffiltext{15}{Finnish Centre of Astronomy with ESO (FINCA), University of Turku, V\"ais\"al\"antie 20 FI-21500 Piikki\"o, Finland}
\altaffiltext{16}{Observat\'orio do Valongo, Universidade Federal de Rio de Janeiro, Ladeira Pedro Ant\^onio, 43, CEP 20080-090, Rio de Janeiro, Brazil}
\altaffiltext{17}{National Radio Astronomy Observatory / NAASC, 520 Edgemont Road, Charlottesville, VA 22903}
\altaffiltext{18}{Department of Physics and Space Sciences, Florida Institute of Technology, 150 W. University Boulevard, Melbourne, FL 32901}
\altaffiltext{19}{Space Telescope Science Institute, 3700 San Martin Drive, Baltimore, MD 21218}

\begin{abstract}
Star-forming regions that are visible at $3.6\mu$m and H$\alpha$ but not in the
{\it u,g,r,i,z} bands of the Sloan Digital Sky survey (SDSS), are measured in five
nearby spiral galaxies to find extinctions averaging $\sim3.8$ mag and stellar
masses averaging $\sim5\times10^4\;M_\odot$. These regions are apparently young
star complexes embedded in dark filamentary shock fronts connected with spiral
arms. The associated cloud masses are $\sim10^7\;M_\odot$. The conditions required
to make such complexes are explored, including gravitational instabilities in
spiral shocked gas and compression of incident clouds. We find that instabilities
are too slow for a complete collapse of the observed spiral filaments, but they
could lead to star formation in the denser parts. Compression of incident clouds
can produce a faster collapse but has difficulty explaining the semi-regular
spacing of some regions along the arms. If gravitational instabilities are
involved, then the condensations have the local Jeans mass. Also in this case, the
near-simultaneous appearance of equally spaced complexes suggests that the dust
lanes, and perhaps the arms too, are relatively young.
\end{abstract}


\section{Introduction}

Direct evidence for spiral wave triggering of star formation has been difficult to
find since density wave shocks were first identified by \cite{fujimoto68}  and
\cite{roberts69}.  Early proposals for triggering involved shock compression of
incoming clouds \citep{woodward76}, diffuse cloud formation in a phase transition
driven by thermal instabilities \citep{shu73}, diffuse cloud formation in the
magnetic instability of Parker (1966; Mouschovias et al. 1974), cloud-cloud
collisions \citep{vogel88,koda06}, and giant cloud formation by gravitational
instabilities in the shocked gas \citep{elmegreen79,kim02}.  Color and age
gradients downstream from the shock suggest star formation triggering
\citep{sheth02,tamburro08,mar09,egusa09,mar11,san11}, but differential compression
of old stars and gas \citep{elm87} and between HI and CO gas phases
\citep{louie13} can produce structures that mimic age gradients. It is sometimes
unclear whether there is a preferential formation of stars in spiral arm gas
compared to interarm gas: some observations show a clear change in molecular cloud
properties or star formation efficiencies inside spiral arms compared to the
interarms \citep{foyle10,hirota11, rebolledo12, sawada12a, sawada12b, koda12}
while other observations do not \citep{foyle10,eden12,eden13}.

Infrared observations of galaxies offer a better view of star formation than has
been possible before. \cite{gros09,gros12} found bright young complexes and
clusters in ten spiral galaxies using J,H,K photometry, and determined the ages,
masses, and extinctions for hundreds of these regions. The extinctions ranged up
to $A_{\rm V}\sim7$ mag for the youngest objects, the masses were up to
$\sim10^6\;M_\odot$, and the ages were typically less than 10 Myr.

Longer-wavelength observations with the {\it Spitzer} space telescope penetrate
dust even better than J, H, K photometry. \cite{prescott07} studied embedded star
formation on 500 pc scales in the {\it Spitzer}-based SINGS survey by comparing
$24\,\mu$m emission with H$\alpha$. They found a median extinction of 1.4 mag in
H$\alpha$, and that 4\% of the $24\,\mu$m sources have H$\alpha$ extinctions
larger than 3 mag.

Here we select the most opaque regions in the optical bands of five nearby spiral
galaxies and search the corresponding {\it Spitzer} $3.6\mu$m images for sources
without optical counterparts. The spatial resolution is $\sim150$ pc or less. We
find that many spiral arm dust filaments contain such sources and determine the
positions, extinctions, and masses of the most prominent objects.  The appearance
of embedded star formation in spiral arm dust filaments suggests that gas collapse
can be rapid in the cores of the filaments.  This is in agreement with recent
models of spiral shocks by \cite{bonnell13}.

\section{Observations and Sample}

The observations reported here are part of the Spitzer Survey of Stellar Structure
in Galaxies \cite[S$^4$G;][]{sheth10}, which is a survey at $3.6\,\mu$m and
$4.5\,\mu$m using the Infrared Array Camera \cite[IRAC;][]{fazio04}. The survey
comprises over 2300 nearby galaxies, including warm mission and archival data that
were processed through the S$^4$G pipeline. The released images have
$0.75^{\prime\prime}$ pixels and a resolution of $1.7^{\prime\prime}$. Five
galaxies (Table \ref{galaxies}) representing different spiral arm classes but all
with about the same luminosity, size, and Hubble type were chosen for inspection
of embedded star formation regions, using optical images from the Sloan Digital
Sky Survey \citep[SDSS;][]{gunn} for comparison. The assumed distances come from
the NASA/IPAC Extragalactic Database. The SDSS database has images in $u, g, r, i,
z$ filters, with $0.396^{\prime\prime}$ pixels. The SDSS images were repixelated
and reoriented to match the {\it Spitzer} images using the {\it wregister} task in
the {\it Image Reduction and Analysis Facility} (IRAF).

The most obvious embedded star-forming complexes were identified by eye on the
$Spitzer$ $3.6\,\mu$m images by blinking these with the SDSS $g$ images in IRAF.
Photometric measurements were done for each identified source using the IRAF
routine {\it imstat} to draw rectangles around the complexes, which are often
elongated. The average background intensity for each region and passband was
measured from linear scans one pixel wide through the center of the region in a
direction perpendicular to the local arm using the IRAF routine {\it pvec}. This
background was then subtracted from the intensities of the regions to produce
source magnitudes.

Ground-based observations with H$\alpha$ narrow-band filters were also compared to
the IR and optical images. NGC~4321, NGC~5055, NGC~5194 H$\alpha$
continuum-subtracted images were obtained from The \textit{Spitzer} Infrared
Nearby Galaxies Survey (SINGS) Legacy
Project\footnote{http://irsa.ipac.caltech.edu/data/SPITZER/SINGS/}. H$\alpha$
continuum-subtracted images of NGC~5248 and NGC~5457 were obtained from
\cite{knapen04}. The NGC~4321, NGC~5055, NGC~5194 H$\alpha$ images have a
resolution of $0.305^{\prime\prime}/$pixel, whereas NGC~5248 and NGC~5457 images
have a resolution of 0.241$^{\prime\prime}/$pixel. All of the information regarding
the observations, data reduction and data products can be found in the SINGS
webpage and in \cite{knapen04}.

The H$\alpha+$NII fluxes were measured for the same regions that were used to
define the $3.6\mu$m sources, but the entire flux emission from these regions was
included, not just the flux confined to the source rectangles (the additional
H$\alpha$ flux from outside the rectangles, estimated to be 10-20\% of the total, is
assumed to be from the embedded sources because HII regions are often larger than
the star complexes that excite them). We have estimated the contribution from the
neighboring NII lines using integrated spectrophotometry of the galaxies from
\cite{mk06} in the case of NGC~4321 and NGC~5194, and an empirical scaling
relation between the NII/H$\alpha$ ratio and M$_{B}$ described in Appendix B of
\cite{kennicutt08} in the case of NGC~5055, NGC~5248 and NGC~5457. The resulting
NII/H$\alpha$ ratios are 0.430 (NGC~4321), 0.486 (NGC~5055), 0.590 (NGC~5194),
0.369 (NGC~5248) and 0.389 (NGC~5457). After correcting for the NII contamination,
the H$\alpha$ fluxes, F(H$\alpha$), are converted to H$\alpha$ luminosities, as $
L({\rm H\alpha}) [{\rm erg/s}]=4\pi D^{2} F({\rm H\alpha})$, with \textit{D} being
the distance to the galaxy in cm (Table 1).

2MASS images at J band ($1.25\mu$m, Skrutskie et al. 2006) of the same galaxies
were measured as well, using the same box sizes as for the $3.6\mu$m sources and
the same background regions for subtraction. The 2MASS pixel sizes are
$2^{\prime\prime}$.  The 2MASS data is used in our application of a technique
proposed by \cite{mentuch10} to estimate the underlying old stellar population. As
discussed more below, this technique assumes that the old stellar population in
the galaxy disk has a flux density at $1.25\mu$m that is typically about twice the
flux density at $3.6\mu$m. Mentuch et al. then showed that the excess $3.6\mu$m
emission above half the $1.25\mu$m emission correlates with H$\alpha$. This gives
an independent measure of the intrinsic H$\alpha$ luminosity of the source,
presumably with less extinction than the direct H$\alpha$ measurement.

Figures 1--5 show the galaxies studied here. The color composites in the top left
panels use SDSS $g$-band images for blue, SDSS $i$-band images for green, and IRAC
$3.6\,\mu$m images for red. They all have $0.75^{\prime\prime}$ pixels and were
combined into color images after first using IRAF {\it wregister} to rotate,
stretch, and align the optical images to match the infrared images. The top right
panel shows an image remade from the {\it u}, {\it r}, and {\it z} SDSS filters
using data from the SDSS webpage (www.sdss.org). The middle left panel shows a
sample region at $3.6\,\mu$m with circles highlighting $3.6\,\mu$m sources that do
not appear in optical bands. These same circles are shown in the middle right
panel on an SDSS $g$-band image. For clarity, these circles are $\sim10\times$
larger than the measurement rectangles. The bottom left image is the $3.6\mu$m
full image with a rectangle indicating the sample region, and the bottom right
image is the sample region again in H$\alpha$.

Comparison of the middle panels in Figures 1--5 shows sources clearly visible at
$3.6\mu$m but invisible at $g$-band. Sometimes part of a $3.6\mu$m source is
visible at $g$ inside the circles, but the measuring rectangle was made on the
part of the $3.6\mu$m source that was not visible at $g$ or at any other SDSS
filter.  Measurement rectangles are typically several arcseconds in length and
width.

The identified sources appear to be embedded star-forming regions in the spiral
arm dust lanes. They are visible at $3.6\,\mu$m because the extinction there is
only $\sim5$\% of the $g$-band extinction \citep{cardelli89}. They are often
visible in H$\alpha$ too, which is strange because that is an optical band, but
H$\alpha$ emission can sometimes be more easily detected than optical starlight,
and some of the H$\alpha$ could be from regions that are between and around the
densest parts of the obscuring clouds. Higher resolution HST images show parts of
these sources to be visible at optical bands, but the HST optical fluxes are still
weak compared to the $3.6\mu$m fluxes because of a generally high obscuration.

The origin of the $3.6\mu$m emission is understood only in general terms. Some of
it is from the Rayleigh-Jeans limit of the photospheric emission from young and
old stars, some of it is from hot dust in young stellar disks and dense cores, and
some is from PAH emission associated with the HII regions. We assume that the PAH
part of the $3.6\mu$m emission is an independent measure of flux from the HII
region. After subtracting the contribution from underlying stars
\citep{mentuch10}, we can then determine the embedded cluster masses from the
required ionizing fluxes. Comparison with the SDSS-band upper limits then gives a
lower limit to the extinction.

The SDSS-band upper limit flux in a dust lane, $F_{\rm up}$, is taken to be the
rms value of the SDSS pixel counts, $\sigma_{\rm px}$, multiplied by the
counts-to-specific flux conversion factor, $C$, divided by the square root of the
number of pixels, $N_{\rm px}^{0.5}$, and multiplied by the student-t value
$t_{0.05}=1.7$ giving 95\% probability that the mean count is less than
$t_{0.05}\sigma_{\rm px}/N_{\rm px}^{0.5}$.  That is, the probability that the
mean flux from the source is less than the measured mean ($<F>\sim0$ here) plus
the increment $t_{0.05}\sigma_{\rm px}C/N_{\rm px}^{0.5}$ is 0.95. Here,
$\sigma_{\rm px}$ comes from the IRAF measurement of the source counts. The value
of $t_{0.05}$ is for a typical number of pixels in a source, $n\sim20$, so the
number of degrees of freedom for the student-t statistic is $n-1$, and the
probability that the true mean $\mu$ is between $<F>-t_{0.05}\sigma_{\rm
px}/N_{\rm px}^{0.5}$ and $<F>+t_{0.05}\sigma_{\rm px}/N_{\rm px}^{0.5}$ is 0.90.
This is the same as saying that
\begin{equation}
F_{\rm up}=t_{0.05}\sigma_{\rm px}C/N_{\rm px}^{0.5}
\end{equation}
with 95\% confidence for $t_{0.05}=1.7$.  The limiting magnitude is obtained from
the limiting source flux using the usual AB conversion, mag$=-2.5\log_{10}F_{\rm
up}-48.60$ for $F_{\rm up}$ in ergs cm$^{-2}$ s$^{-1}$ Hz$^{-1}$.

Table \ref{results} lists the positions of the embedded sources, the number of
pixels of $0.75^{\prime\prime}$ size in the measured rectangles, their fluxes,
$S_{\nu}$, at $3.6\,\mu$m in $\mu$Jy, AB absolute magnitudes in IRAC $3.6\,\mu$m,
AB absolute magnitude upper limits in the SDSS bands, and the H$\alpha$
luminosities measured from ground-based observations.

\section{Method}

Without optical detections of the embedded sources and only IRAC observations at
$1.7^{\prime\prime}$ resolution from {\it Spitzer} and $1.25\mu$m observations at
$2^{\prime\prime}$ resolution from 2MASS, we cannot uniquely determine the source
ages, masses and extinctions. However, the sources are all embedded in spiral arm
dust lanes and likely to be young \cite[e.g.,][]{scheepmaker09}, as also implied
by the H$\alpha$, so we searched over a range of ages, and for each age determined
the cluster mass and extinction required to give the observed $3.6\mu$m and
$1.25\mu$m fluxes using the conversion to intrinsic H$\alpha$ in \cite{mentuch10}:
\begin{equation}
I_{\rm IR}(H\alpha)=\left(I[3.6\mu{\rm m}]-0.5
I[1.25\mu{\rm m}]\right)/0.043
\end{equation}
for $I$ in units of MJy/ster.  That is, $I(3.6\mu{\rm m})$ and $I(1.25\mu{\rm m})$
in MJy/ster were measured from the background subtracted specific fluxes in each
source (in units of $\mu$Jy from the image source counts), converted to MJy and
divided by the area of the source in ster. The resulting $I_{\rm IR}(H\alpha)$ was
then multiplied by a typical H$\alpha$ filter width in frequency units, $\Delta
\nu$, which comes from the filter width in wavelength units, $\Delta \lambda=75
$\AA, converted to frequency with the equation $\Delta
\nu=c\Delta\lambda/\lambda^2$. This gives the source H$\alpha$ flux in units of
erg cm$^{-2}$ s$^{-1}$.

The intrinsic H$\alpha$ flux derived in this way from the IR, and the directly
measured H$\alpha$ flux from ground-based observations, were converted to
H$\alpha$ luminosities $L_{{\rm H}\alpha}$ by multiplying by $4\pi D^2$ for
distance $D$, and then to observed ionization rates $S_{\rm obs}$ using the
equation $S_{\rm obs}=7.3\times10^{11}L_{{\rm H}\alpha}$ s$^{-1}$
\citep{kennicutt98}. These observed ionization rates were compared to stellar
evolution models to determine intrinsic source properties.

For the stellar models, we used tables in \cite{bruzual03} for AB magnitude versus
time in {\it u,g,r,i}, and {\it z} bands, which are given for a unit of solar mass
in an initial cluster with a fully sampled Chabrier IMF and solar abundances. We
also used the tables for Lyman continuum ionization rate versus time per unit
initial solar mass, and residual stellar mass versus time per unit initial solar
mass. The optical magnitudes were converted to luminosities per unit initial mass,
and these along with the Lyman continuum rate and the residual stellar mass, were
integrated over time up to the assumed age using a constant star formation rate.
This integral over the model Lyman continuum rate per unit initial mass will be
denoted by $S_{\rm m}(t)$, and the integral over the model residual mass per unit
initial mass will be denoted by $M_{\rm m}(t)$.

The ratio of the observed ionization rate $S_{\rm obs}$ to the model rate $S_{\rm
m}$ gives the correction factor to the model integrated mass, from which the
required cluster mass can be derived:
\begin{equation}
M_{\rm cluster} = \left({{S_{\rm obs}}\over{S_{\rm m}(t)}}\right)M_{\rm m}(t)
\end{equation}
The integrated luminosities in each SDSS band were found from the same ratio,
e.g., $L_{\rm u}=(S_{\rm obs}/S_{\rm m}[t])L_{\rm m,u}(t)$ for {\it u}-band, and
converted back to apparent magnitude. The resulting model magnitudes in all five
SDSS bands were then subtracted from the observed magnitude limits in each band to
get the minimum extinction in each band.  These were all converted to {\it g}-band
extinction using the extinction curve in \cite{cardelli89}, and the results were
averaged together. This extinction curve should be more appropriate for embedded
sources than the gray type of extinction that comes from a partially transparent
screen \citep{gros12}. The different limiting g-converted extinctions varied among
the passbands by an average of 0.5 mag, presumably because the SDSS limits were
not the same depth relative to the spectral energy distribution of the source.

\section{Results}
\label{results}

Table 3 gives the derived properties of the embedded sources listed in Table 2 for
assumed ages of 3 Myr. The masses and average {\it g}-band-equivalent extinctions
for the ground-based H$\alpha$ measurements  are in columns 3 and 4, the masses
and average {\it g}-band-equivalent extinctions for the $3.6\mu$m measurements are
in columns 5 and 6, and the extinction needed in the HII region, measured at {\it
g}-band, to produce the observed H$\alpha$ fluxes given the star forming regions
measured at $3.6\mu$m, are in column 7. The sources with lower limits have upper
limits for the directly observed H$\alpha$ fluxes (more on this below and in the
footnote to table 3). All sources are observed at $3.6\mu$m.

The derived cluster masses increase with assumed age in the range from 1 to 10 Myr
by about a factor of 2.5 for all sources (not shown). This is small enough for the
likely age range that we consider here only masses for an assumed age of 3 Myr, as
tabulated, with an uncertainty of a factor of 1.6 either way from unknown age
effects. The mass differences obtained using the two methods, direct H$\alpha$
observations and H$\alpha$ inferred from $3.6\mu$m and $1.25\mu$m emissions,
typically differed by a factor of 24 with the IR-based masses being larger. This
difference amounts to an excess extinction of about 3.5 magnitudes for the direct
H$\alpha$ compared to the IR-derived H$\alpha$.

Table 4 gives the embedded source masses and extinctions for the two types of
measurements and the equivalent $g$-band extinction in the HII region to give
H$\alpha$ consistent with $3.6\mu$m and $1.25\mu$m, averaged over all sources in
each galaxy. The masses measured by direct H$\alpha$ observations average
$2.4\times10^3\;M_\odot$, and the masses measured by $3.6\mu$m and $1.25\mu$m
observations and converted to H$\alpha$ average $5.0\times10^4\;M_\odot$. The
lower limits to the extinctions provided by the upper limits to the SDSS fluxes
and the ground-based H$\alpha$ fluxes (column 3) are about zero, meaning that the
H$\alpha$ measurements are consistent with the optical broad-band measurements in
being highly extincted.  In other words, given the stellar masses required to
produce the (highly extincted) H$\alpha$ fluxes that were directly observed, the
expected $u,g,r,i,z$ fluxes are comparable to the upper limits within a few tenths
of a magnitude (the error bars in column 3 of Table 4).  The lower limits to the
extinctions derived from the $3.6\mu$m fluxes in comparison to the SDSS bands
(column 5) are, on average, $\sim3.8$ mag for all galaxies. This is comparable to
the extinction in the HII region (column 6) because the HII emission is extincted
by about the same amount as the SDSS bands.

We conclude that these dusty regions are obscuring young stellar associations with
an average mass of $5.0\times10^4\;M_\odot$ for an age of about 3 Myr. The average
mass would be $4.2\times10^4\;M_\odot$ for an age of 1 Myr and
$1.3\times10^5\;M_\odot$ for an age of 10 Myr.  The average extinction to these
regions in $g$-band is $\sim3.8$ magnitudes.

\section{Discussion}

We consider possible star formation mechanisms as summarized in the introduction.
Two of these mechanisms, the Parker and thermal instabilities, involve diffuse gas
without self-gravity. While they may contribute to interstellar structure, they
are not involved with rapid star formation because that requires self-gravity and
collapsing cloud cores. The other mechanisms, gravitational instabilities and
cloud impacts or collisions, involve strong self-gravity from an early stage, the
first because whole dust lanes can be unstable to fragmentation along their
lengths, and the second because collisions increase a cloud's density and that can
make the cloud gravitationally unstable.

The observation of bright embedded infrared sources in galactic spiral shocks
implies that massive star formation begins quickly in the shocked gas -- faster
than the time delay usually inferred from the displacement between dust lanes or
CO emission and visible HII regions \citep[e.g.,][]{egusa09}. The observed mass of
$\sim10^4-10^5\;M_\odot$ is comparable to the masses of more visible clusters and
star-forming regions in normal galaxies.

In what follows, we consider two possible formation mechanisms for the observed
star complexes. For both, the dust lanes are assumed to be shocks in spiral
density waves. This is consistent with the streaming motions that are often
observed in CO observations of these regions \citep[e.g.,][]{shetty07}. Most of
the molecular clouds in the Milky Way spiral arms \citep{dame01} seem related to
spiral shocks too, considering their velocities and positions in the global spiral
pattern \citep{bissantz03}. Their concentration in spiral arms is apparent in the
face-on view of the Milky Way shown by \cite{englmaier11}. Evidently, these local
CO clouds would resemble dust lanes if they were viewed from outside the Milky
Way.

The two mechanisms considered are (1) gravitational collapse of gas that
accumulates behind the shock front in a spiral arm, and (2) compression and
collapse of pre-existing clouds that hit these shock fronts.  In the following
discussion, we suggest that gravitational collapse of spiral arm shock fronts is
too slow on kpc scales to account for the rapid onset of star formation that
appears to be required for these embedded sources.  The densest parts of the
shocked gas could have collapsed into stars, however.  Compression of interarm
clouds that hit previously shocked gas could trigger the obscured sources faster.
Additional considerations are needed to determine which mechanisms apply.

\subsection{Gravitational Collapse of Spiral Arm Shock Fronts}

Table \ref{averages} suggests that a typical value of optical extinction to an
embedded dust lane source is $\sim3.8$ mag. An embedded source with $A_V=3.8$ mag
of extinction may be inside a dust cloud that has $\sim8$ mag of extinction all
the way through it if the source is centered.  This makes the total column density
through the cloud $\Sigma_{\rm gas}\sim160\;M_\odot$ pc$^{-2}$ for
$N(HI)=5.8\times10^{21}E_{\rm B-V}$ \citep{bohlin78}, ratio of total-to-selective
extinction $R_{\rm V}=3.1$ \citep{draine03}, and mean atomic weight 1.36. A
typical dust lane width is $\sim4$ pixels in Figures 1--5 (1
px$=0.75^{\prime\prime}$), which is $W\sim150$ pc at a distance of 10 Mpc. The
dust lane linear density is the product of these, $\mu=\Sigma_{\rm
gas}W\sim2.4\times10^4\;M_\odot$ pc$^{-1}$, or $1.5\times10^{19}$ g cm$^{-1}$. The
critical line density for collapse of a filament is $\mu_{\rm c}=2\sigma^2/G$ for
1-dimensional velocity dispersion $\sigma$ \citep{inutsuka92,inutsuka97}. If
$\sigma\sim7$ km s$^{-1}$ in the shock, then $\mu_c$ is comparable to the observed
line density. This suggests that spiral arm shock fronts are unstable to collapse
along their lengths \citep{elmegreen79,kim02,kim07,renaud13}.

The result of this collapse would be one or more giant cloud complexes with masses
equal to $\sim S\mu W$ for spacing $SW$ between the complexes along the arm
\citep{inutsuka97}. Taking $S\sim4$ from the figures here and from
\cite{inutsuka97}, the resulting cloud mass is $\sim1.4\times10^7\;M_\odot$. If we
consider a typical efficiency of 2\% for the star-forming complex, then a stellar
mass of $\sim3\times10^5\;M_\odot$ results, large enough to account for what we
observe.

The key to star formation in spiral arm dust lanes is that it has to happen
quickly, before the spiral arm passes and the shocked gas decompresses in the
interarm. This is a strong constraint if the spiral wave is steady. Not all
compressed regions will be part of a steady flow, however. If a dust lane is
sufficiently self-gravitating, then it can remain intact as a giant cloud complex
even when it enters the interarm region, making its lifetime longer than the arm
crossing time. Such shocks would not be steady but would cycle between gas build
up, collapse, and dispersal.  If the lifetime of the compressed state is long
enough, then interarm protrusions and spurs result
\citep{wada04,lavigne06,shetty06,dobbs06,lee12,renaud13} with star formation
lingering in the interarm regions.

\cite{egusa11} observed M51 with 30 pc resolution and could see cloud build-up
from small interarm CO clouds to giant molecular associations as the gas flowed
into and through the spiral arms. With 780 pc resolution, the excitation state of
M51 CO clouds changes from low in the interarm regions to high in the arms,
corresponding to an increase in density and/or temperature with increasing star
formation activity \citep{koda12}. There is also an indication in M51 of dense CO
gas just before prominent spiral arm star formation; the map in Figure 6 of
\cite{koda12} shows regions with highly excited CO and low $24\,\mu$m emission.
These regions appear to be centered on dust lanes, although their exact spatial
distribution is uncertain because the 780 pc resolution scale of the map is much
larger than a dust lane. These regions avoid the parts of the dust lanes where we
identify the brightest embedded sources, but that is expected from the lack of
$24\,\mu$m emission in the \cite{egusa11} sources. Similarly, \cite{hirota11}
observed molecular clouds in IC 342 turn from weakly self-gravitating in the
interarm regions to strongly self-gravitating and star-forming in the arms.

We now consider when the collapse of density wave-shocked gas is fast enough
to make stars in the steady state case. This requires a comparison between
the collapse time of this gas and its dwelling time in the arm.

The dwelling time for gas in a dust lane is the arm-to-arm time multiplied by
the ratio of the gas mass inside the dust lane within some radial interval to
the total gas mass from one arm to the next in that radial interval. This
relationship is from the continuity equation. Because of streaming motion
along the arm in the dust lane, this dwelling time is longer than the purely
geometric time obtained from the ratio of the angular thickness of the dust
lane measured from the galaxy center divided by the angular rate of the
spiral pattern relative to the gas.

The ratio of the dust lane mass to the total arm-to-arm gas mass in a radial
interval equals the ratio of the dust lane column density to the average column
density, multiplied by the relative angular extent of the dust lane. In the
star-forming part of the main disk of most spiral galaxies, the {\it V}-band
extinction averaged in azimuth is typically about 1 magnitude. Recall that the
column density for general extinction $A_{\rm V}$ equals $20A_{\rm V}\;M_\odot$
pc$^{-2}$ in units of surface density, and an average value of $20\;M_\odot$
pc$^{-2}$ for spiral galaxy disks is typical inside the molecular zone
\citep{bigiel08}. For a general case like this, the ratio of column
densities is then approximately $A_{\rm V,dl}/(1\;{\rm mag})$ for $A_{\rm V,dl}$
in the dust lane. The arm-to-arm distance is $\pi R$ for a two-arm spiral at
radius $R$. Taking $R=4$ kpc as representative, the relative dust lane thickness
is $W/\pi R\sim 1$\% with $W=150$ pc from above. Thus the fraction of the gas mass
in the dust lane, which is also the fraction of the arm-to-arm time in the dust
lane, is approximately $0.01A_{\rm V,dl}$. The arm-to-arm time is $\sim120$ Myr
for a position at $R\sim4$ kpc that is halfway to corotation where the galaxy
rotation speed is $\sim200$ km s$^{-1}$. Then the dwelling time for gas in the
dust lane is $\sim1.2A_{V,dl}$ Myr. In general, the dust lane flow-through time is
$(W/V_{\rm rel})\times(A_{\rm V,dl}/A_{\rm V,ave})$ for dust lane width $W$,
azimuthal speed $V_{\rm rel}$ of the spiral arm relative to the gas, and average
extinction through the disk at the radius of the dust lane, $A_{\rm V,ave}$. This
result does not depend on $R$ explicitly. If $W$ is in pc and $V_{\rm rel} $ is in
km s$^{-1}$, then the unit of this flow-through time is approximately in Myr.

The fraction of the time that gas spends in the vicinity of a spiral arm can be
fairly high for a strong arm. For example, a CO map of the galaxy shown in Figure
4, NGC 5248 \citep{kuno07}, has two spiral arms midway out with $\sim12$ K km
s$^{-1}$ integrated intensities that occupy about 20\%-25\% of the azimuthal
distance to the next arm, and it shows interarm regions with less than 3 K km
s$^{-1}$. Thus the arms contain about half of the gas at that radius
($(12/3)*0.25=0.5$), and the arm dwelling time is half of the arm-to-arm time if
most of the gas is molecular. This fraction would be lower if the $X$-factor that
converts CO to H$_2$ is lower in the arms than in the interarms because of greater
CO temperatures in the arms. Using $V_{\rm rel}\sim100$ km s$^{-1}$, which is half
of the rotation speed of NGC 5248 \citep{nishiyama01}, assuming this position is
half the distance to co-rotation, and assuming a galactocentric distance of 1
arcmin $\sim4$ kpc, the arm-to-arm time is 126 Myr and the arm dwelling time could
be about 60-70 Myr. In the spiral arms of M51 \citep{koda11}, the dust lanes
appear as thin ribbons of CO emission nearly resolved by CARMA. They occupy about
10\% of the azimuthal angle and have 5 to 10 times the average interarm intensity
in CO. Thus even the thin CO arms in M51 contain 30\% to 50\% of the molecular
mass at that radius, and because most of the gas in this galaxy is molecular
\citep{koda09}, this is also approximately the fraction of the arm-to-arm time
spent in the dust lanes (unless the $X$-factor varies).


The dwelling time in a spiral arm has to be compared with the timescale for
self-gravitational collapse. The latter is about $S(G\rho)^{-1/2}$ for
length-to-width ratio $S$ along the dust lane and for density $\rho$. The factor
$S$ comes from the average equivalent density $\rho_{\rm sph}$ in a sphere
containing the fragment, considering that the collapse time is $\sim(G\rho_{\rm
sph})^{1/2}$ for that average.  In agreement with this, \cite{inutsuka97} show one
case of a filament with a collapse to a singularity in $12.3/(4\pi G\rho)^{1/2}$,
which is $3.5/(G\rho)^{1/2}$, giving $S\sim3.5$ in their case \cite[see
also][]{pon12}. The average dust lane density is the extinction through the disk,
$A_{\rm V,dl}$, converted into grams cm$^{-2}$, divided by the line-of-sight
thickness, which is $H\sim70$ pc for a typical molecular disk. For $H$ in parsecs,
this density is $1.4\times10^{-21}A_{\rm V,dl}/H$ g cm$^{-3}$, and the collapse
time is $3.3S(H/A_{\rm V,dl})^{1/2}$ Myr, or some $\sim10S$ Myr if $H\sim70$ pc
and $A_{\rm V,dl}\sim8$ mag (i.e., using $A_{\rm V,dl}\sim2<A_{\rm g}>$ or
$2\Delta A_{\rm g}$ from table 4).

For gravitational collapse in a dust lane, the collapse time has to be less than
the flow-through time, which means $A_{\rm V,dl}^{1.5} > 3.3SH^{1/2}A_{\rm
V,ave}V_{\rm rel}/W$ or $A_{\rm V,dl}>18$ mag for $S=4$, $H=70$ pc, $A_{\rm
V,ave}=1$ mag, $V_{\rm rel}=100$ km s$^{-1}$, and $W=150$ pc. This result implies
that sufficiently dense spiral shocks, i.e., those with extinctions through the
disk greater than $\sim18$ mag for these numbers, should have time to collapse
along their lengths into new cloud complexes.  Because self-gravity is important
from the start in such an instability, the cloud complexes that form will also be
strongly self-gravitating, and should form stars at an accelerating pace while
they collapse further.

Recall that our sources imply $A_{\rm A,dl}\sim8$ mag on average, considering we
see the foreground extinction equal to half of this value.  This extinction is
slightly less than the critical value of 18 mag derived above, suggesting that the
dust lanes we observe are only marginally unstable to collapse in the short time
they have. If the line-of-sight cloud thicknesses were much smaller than the
assumed $H\sim70$ pc, such as the size of a giant molecular cloud, $\sim10$ pc,
then $A_{\rm V,dl}\sim9$ pc, and this would be closer to our observations. In
either case, the observed dust lanes do not appear to be strongly self-gravitating
and in the process of rapid collapse.  They could be forming stars only in the
initially dense regions that were most unstable before they entered the spiral
arm. Such partial collapse is consistent with the spiral shock models by
\cite{bonnell13} and with the growing evidence for star formation in marginally
bound or unbound molecular clouds \cite[i.e., but which are still unstable to
collapse in their cores;][]{dobbs11}.

\cite{kimostriker02} ran simulations showing the steady accumulation of shocked
gas in spiral arms, with dust lane density perturbations building up over several
rotation periods from initially small amplitudes. Eventually the irregularities
caused by gaseous self-gravity led to the formation of giant, self-gravitating
clouds.  This is essentially the process we have in mind here for the formation of
the observed embedded sources.

\cite{elm12} showed examples of beads-on-a-string of star formation along spiral
arms of all types, ranging from two-arm grand design spirals to multiple, long-arm
spirals, to flocculent spirals (see also Elmegreen \& Elmegreen 1983).
Beads-on-a-string patterns were studied in detail by \cite{elm06} using IRAC
images like those considered here. In many images from the {\it Galaxy Evolution
Explorer} satellite ({\it GALEX}; Martin et al. 2005), one gets the impression
that most star formation is in complexes that are strung out along spiral arms in
a semi-regular fashion. Interstellar magnetic fields may be involved in this
regularity \citep{efremov}. Evidently, there is a close resemblance between spiral
arm star formation on kpc scales and filamentary star formation \citep{andre}
inside molecular clouds on parsec scales.

\subsection{Pressurized Gravitational Collapse of Incident Massive Cloud Complexes}

Compression and collapse of an incident interarm cloud can be faster than the
collapse time calculated above for an initially uniform dust lane. The compression
time $t_{\rm comp}$ is the cloud size $R_{\rm cloud}$ divided by the relative
perpendicular speed between the cloud and the spiral arm, $V_{\rm perp}$. For a
relative azimuthal speed between the cloud and the arm, $V_{\rm rel}=100$ km
s$^{-1}$ as above, the perpendicular speed is $V_{\rm rel}\sin i\sim30$ km
s$^{-1}$ when the spiral arm pitch angle is $i=20^\circ$. Then $t_{\rm comp}\sim3$
Myr for $R_{\rm cloud}\sim100$ pc. Collapse would follow compression quickly
because the compressed cloud density is high, approximately equal to the product
of the interarm cloud density, $20A_{\rm V,ia}/H$ (in $M_\odot/$pc$^{3}$ for an
interarm cloud extinction $A_{\rm V,ia}$ and line-of-sight thickness $H$ in pc)
multiplied by the square of the ratio of the perpendicular compression speed to
the cloud thermal speed, $\left(V_{\rm perp}/v_{\rm th}\right)^2$. For $A_{\rm
V,ia}\sim1$ mag, $H=70$ pc, $V_{\rm perp}=30$ km s$^{-1}$, and $v_{\rm th}=3$ km
s$^{-1}$, the compressed cloud density is $n_{\rm comp}\sim850$ atoms cm$^{-3}$
and the cloud collapse time, measured as $(G\rho_{\rm comp})^{-1/2}$, is $\sim2.8$
Myr.  If such a density extends for $H=70$ pc along the line of sight, then the
extinction through the disk would be $A_{\rm V}\sim100$ mag and the extinction to
an embedded source $\sim50$ mag. This is significantly larger than what we observe
here for the clouds on $W\sim150$ pc scales, but could apply in their cores.

The original picture of cloud compression in spiral shocks was based on
simulations by \cite{woodward76} of a small diffuse cloud engulfed by a rapidly
moving, infinitely-extended shock front. The cloud is compressed by the post-shock
gas and forms a cometary structure, with the head collapsing into stars. Many
simulations have confirmed this basic process. Spiral arms do not typically show
cometary clouds and the dust lanes surrounding the embedded sources discussed here
are not cometary. Still, the basic model of cloud compression upon impact with an
existing spiral arm dust lane (from a previous shock) seems generally valid.

The interarm clouds in Figures 1--5 are spiral-like with large cloud complexes
distributed along their lengths. This interarm structure is also evident from
well-resolved maps of IR \citep{block97} and CO \citep{koda11}, as in the case of
M51. Such shapes are remnants of dust lane structures, secondary shocks, and spurs
from the previous arms \citep{lavigne06,shetty06,dobbs06}. Interarm cloud
complexes can be moderately massive: $M_{\rm cloud}=6\times10^5\;M_\odot$ if
$R_{\rm cloud}\sim100$ pc and $A_{\rm V,ia}=1$ mag. The most opaque interarm
clouds may contain $\sim10^7\;M_\odot$ and have a substantial molecular fraction
\citep[e.g.,][]{koda11}.

The observed masses of the largest embedded complexes found here are consistent
with interarm cloud compression. The pre-collision clouds would have to be
$\sim50$ times more massive than the observed stellar masses, considering a star
formation efficiency on $\sim100$ pc scales of 2\%. To form a $\sim10^5\;M_\odot$
star complex inside a dust lane (Table 4), the previous interarm cloud would have
to contain $5\times10^6\;M_\odot$. Such a cloud, if spherical, would be $\sim280$
pc in diameter if its average extinction is $A_{\rm V,ia}=1$ mag. With such a size
in two dimensions, the cloud would form a $\sim280$ pc long dust filament when it
hits the spiral arm, and then the collapse to star formation would be like the
dust lane collapse discussed above, with motions along its length and a factor
$S\sim4$ geometric dilution of the collapse time.

\subsection{Regular spacing of star formation along spiral arms}

Regular spacing of large star complexes along spiral arms, as occasionally seen
in Figures 1--5 and in other galaxies, implies that an instability might be
involved and we are seeing the fastest growing wavelength of that instability
\citep[e.g.,][]{wada04,shetty06,lee12,renaud13}. The presence of several young
sources in this string implies that the instability operates all along the
length of the pattern with a nearly uniform starting time, to within some
variation equal to the instability time or the arm dwelling time. Such
simultaneity requires that the dust lanes with regular star formation evolved
within this time interval, that is, they built up, fragmented, and turned into
stars along the length of the regular pattern within several tens of Myrs. After
this, another dust lane would be expected to build up to replace it using fresh
interarm gas. This would be the case for a continuously moving spiral arm that
accumulates gas, dumps it into star complexes and then accumulates more gas.
Alternatively, the whole stellar and gaseous arm could have formed recently as a
result of a spiral instability in the disk \citep{baba13}. Then the arm
structure itself would be fairly young (although the stellar populations in it
would be mostly old) and it could just now be forming its first string of star
formation, i.e., within the last $\sim30$ Myr. In both cases, the instability
would be slow compared to the dynamics of star formation, so the OB associations
that form would be in various stages of obscuration and break-out, ranging from
embedded for the youngest to fully exposed with adjacent cloud debris for the
oldest in the same arm segment.

In cases without a regular spacing along spiral arms, a similar time sequence may
occur with spiral-shocked gas followed by partial cloud collapse, star formation,
and cloud dispersal, but with more of a distributed starting time. Then there will
be different phases in this process at different positions along the arm. There
could also be other processes acting, such as direct compression of interarm
clouds, which would be randomly placed with respect to the regular structures.

\section{Conclusions}

A search for emission at $3.6\mu$m that has no associated optical counterpart in
SDSS images and only weak emission at H$\alpha$ revealed 73 sources that are
likely to be young star formation regions embedded in molecular clouds. The clouds
appear as dark dust lanes in the spiral arms. The intrinsic H$\alpha$ emission of
each source was estimated from the $3.6\mu$m and $1.25\mu$m emissions using the
method of \cite{mentuch10}, and the luminosity of the associated stars was
determined after assuming an age in the range of 1 to 10 Myr.  This stellar
luminosity, combined with upper limits from the SDSS images, gave lower limits to
the extinction and stellar mass. The observed H$\alpha$, compared with the
H$\alpha$ estimated from $3.6\mu$m and $1.25\mu$m, gave a consistent value for the
extinction, which is apparently near the estimated lower limit (or else the
H$\alpha$ would not be observed either).

The average mass of an embedded source was found to be $5\times10^4\;M_\odot$ and
the average extinction at $g$-band is $3.8$ mag. These sources are often parts of
regular strings of star formation with additional embedded sources and other
sources in various stages of lower obscuration.  Two mechanisms for triggering
this star formation were considered: gravitational collapse of spiral density
wave-shocked gas and gravitational collapse of compressed incident clouds. The
spiral shock model suggests that the clouds we are observing are not strongly
self-gravitating because they have not had enough time after their formation in
the spiral arm. Still, collapse should occur in dense cores of these clouds even
if the overall cloud complexes are weakly bound. This is consistent with recent
ideas that star formation occurs in the dense cores of generally unbound molecular
clouds. The incident cloud model is also consistent with our observations if,
again, the shock compression and collapse to star formation occurs primarily in
unseen dense cores.

The regular distribution of star formation along spiral arms, including the
embedded sources as well as the exposed sources, suggest that a mild instability
is operating to form weakly bound molecular clouds simultaneously along kpc
lengths. Simultaneity on these scales requires a uniform initial starting time for
the instability, to within the unstable growth time or the flow-through time. In
that case the arm could be relatively young, or the shock front in the arm could
be relatively new.

B.G.E. is grateful to E. Mentuch for clarifications about the analysis method that
her team proposed. E.A. and A.B. thank the CNES for financial support. We
acknowledge financial support to the DAGAL network from the People Programme
(Marie Curie Actions) of the European Union's Seventh Framework Programme
FP7/2007-2013/ under REA grant agreement number PITN-GA-2011-289313. This research
made use of the NASA/IPAC Infrared Science Archive, which is operated by the Jet
Propulsion Laboratory, California Institute of Technology, under contract with the
National Aeronautics and Space Administration. This work is based in part on
observations made with the Spitzer Space Telescope, which is operated by the Jet
Propulsion Laboratory, California Institute of Technology under a contract with
NASA. This research made use of the NASA/IPAC Extragalactic Database which is
operated by JPL, Caltech, under contract with NASA.  This research made use of
data products from the Two Micron All Sky Survey, which is a joint project of the
University of Massachusetts and the Infrared Processing and Analysis
Center/California Institute of Technology, funded by the National Aeronautics and
Space Administration and the National Science Foundation. Funding for the SDSS and
SDSS-II has been provided by the Alfred P. Sloan Foundation, the Participating
Institutions, the National Science Foundation, the U.S. Department of Energy, the
National Aeronautics and Space Administration, the Japanese Monbukagakusho, the
Max Planck Society, and the Higher Education Funding Council for England. The SDSS
Web Site is http://www.sdss.org/. The SDSS is managed by the Astrophysical
Research Consortium for the Participating Institutions. The Participating
Institutions are the American Museum of Natural History, Astrophysical Institute
Potsdam, University of Basel, University of Cambridge, Case Western Reserve
University, University of Chicago, Drexel University, Fermilab, the Institute for
Advanced Study, the Japan Participation Group, Johns Hopkins University, the Joint
Institute for Nuclear Astrophysics, the Kavli Institute for Particle Astrophysics
and Cosmology, the Korean Scientist Group, the Chinese Academy of Sciences
(LAMOST), Los Alamos National Laboratory, the Max-Planck-Institute for Astronomy
(MPIA), the Max-Planck-Institute for Astrophysics (MPA), New Mexico State
University, Ohio State University, University of Pittsburgh, University of
Portsmouth, Princeton University, the United States Naval Observatory, and the
University of Washington.

\clearpage

\begin{deluxetable}{lcccccccc}
\tabletypesize{\scriptsize}
\tablecaption{Galaxies Observed for Embedded Sources}
\tablehead{
\colhead{Galaxy} &
\colhead{Type} &
\colhead{Radius (arcsec)} &
\colhead{Inc. (deg)} &
\colhead{D (Mpc)}&
\colhead{pixel (pc)\tablenotemark{a}}&
\colhead{$B_{\rm T,0}$ (mag)} &
\colhead{$M_{\rm B}$ (mag)} &
\colhead{$M_{3.6\,u{\rm m}}$ (mag)\tablenotemark{b}}
}\startdata

NGC 4321 & SXS4	 & 222 & 32 & 20.9 & 76 &  9.98 & -21.62 & -22.57\\
NGC 5055 & SAT4	 & 377 & 55	& 7.54 & 27 &  9.03 & -20.36 & -21.37\\
NGC 5194 & SAS4P & 336 & 52	& 7.54 & 27 &  8.67 & -20.72 & -21.83\\
NGC 5248 & SXT4	 & 184 & 44	& 15.4 & 56 & 10.63 & -20.31 & -21.23\\
NGC 5457 & SXT6  & 865 & 21 & 4.94 & 18 &  8.21 & -20.26 & -20.65
\enddata
\tablenotetext{a}{The resolution of the images is $1.7^{\prime\prime}$, which is
2.26 times the resolution of the pixels ($0.75^{\prime\prime}$) given here.}
\tablenotetext{a}{Mu\~noz-Mateos et al. (2013, in prep).}
\label{galaxies}
\end{deluxetable}

\clearpage

\begin{deluxetable}{lccccccccccc}
\tabletypesize{\scriptsize}
\tablecaption{Measurements of Embedded Sources}
\tablehead{
\colhead{Galaxy} &&&&&&&&&&&\\
\colhead{Region} &
\colhead{RA} &
\colhead{Dec} &
\colhead{$N_{\rm px}$} &
\colhead{$S_{\nu}(3.6\,\mu{\rm m})$} &
\colhead{$M_{\rm 3.6}$} &
\colhead{$u$} &
\colhead{$g$} &
\colhead{$r$} &
\colhead{$i$} &
\colhead{$z$} &
\colhead{$\log_{10}L_{H\alpha}$}\tablenotemark{a}\\
\colhead{} & \colhead{} & \colhead{} & \colhead{} & \colhead{($\mu$Jy)} &
\colhead{} & \colhead{} & \colhead{} & \colhead{} & \colhead{} & \colhead{} &
\colhead{(erg s$^{-1}$)} }
\startdata
NGC 4321 &&&&&&&&&&&\\
 1 & 12:22:50.498 & 15:50:24.03 & 30 &  282 & -13.83 &  -6.09 &  -6.44 &  -6.60 &  -6.80 &  -7.05 &   38.47\\
 2 & 12:22:50.238 & 15:50:19.91 & 36 &  240 & -13.65 &  -6.37 &  -6.74 &  -7.13 &  -7.02 &  -7.36 &   38.66\\
 3 & 12:22:48.705 & 15:50:14.28 & 16 &  220 & -13.56 &  -5.88 &  -6.04 &  -6.54 &  -6.74 &  -7.20 &   38.19\\
 4 & 12:22:48.471 & 15:50:11.66 & 24 &  254 & -13.71 &  -6.23 &  -6.77 &  -7.21 &  -7.30 &  -7.80 & $<  37.73$ \\
 5 & 12:22:48.263 & 15:50:09.03 & 25 &  297 & -13.88 &  -6.75 &  -7.20 &  -7.35 &  -7.31 &  -7.80 &   38.09\\
 6 & 12:22:48.238 & 15:49:47.28 & 20 &  219 & -13.55 &  -5.93 &  -5.84 &  -6.80 &  -6.49 &  -6.97 &   38.14\\
 7 & 12:22:47.614 & 15:49:48.78 & 16 &  166 & -13.25 &  -6.35 &  -6.24 &  -7.13 &  -6.79 &  -7.44 &   38.34\\
 8 & 12:22:46.575 & 15:48:55.15 &  9 &   32 & -11.48 &  -5.58 &  -5.91 &  -5.85 &  -5.93 &  -7.25 & $<  36.75$ \\
 9 & 12:22:46.445 & 15:48:50.65 &  6 &   32 & -11.45 &  -6.98 &  -5.23 &  -5.62 &  -6.63 &  -7.74 & $<  36.94$ \\
10 & 12:22:48.654 & 15:48:04.53 & 16 &  102 & -12.72 &  -5.63 &  -6.26 &  -7.10 &  -7.43 &  -7.39 & $<  37.43$ \\
11 & 12:22:56.241 & 15:48:10.16 & 48 & 1445 & -15.60 &  -6.20 &  -6.30 &  -7.20 &  -6.82 &  -7.10 &   38.74\\
12 & 12:22:51.642 & 15:49:33.41 & 20 &  596 & -14.64 &  -6.98 &  -7.13 &  -7.54 &  -7.27 &  -7.56 &   38.57\\
13 & 12:22:52.603 & 15:49:39.04 & 30 &  364 & -14.10 &  -5.72 &  -6.14 &  -6.68 &  -6.79 &  -7.14 &   38.38\\
14 & 12:22:51.797 & 15:50:05.66 &  9 &   60 & -12.14 &  -6.50 &  -6.41 &  -7.01 &  -7.08 &  -7.93 &   37.73\\
15 & 12:22:54.656 & 15:49:33.04 &  9 &  383 & -14.16 &  -6.37 &  -7.41 &  -8.07 &  -8.54 &  -9.13 &   38.40\\
NGC 5055 &&&&&&&&&&&\\
 1 & 13:15:50.565 & 42:01:16.91 &  6 &   63 &  -9.98 &  -7.05 &  -6.38 &  -7.38 &  -7.94 &  -9.15 & $<  37.11$ \\
 2 & 13:15:49.657 & 42:01:12.78 & 12 &  187 & -11.17 &  -7.95 &  -8.40 &  -9.09 &  -9.31 &  -9.99 &   37.26\\
 3 & 13:15:48.512 & 42:01:10.91 & 36 &  651 & -12.52 &  -5.52 &  -5.39 &  -5.95 &  -6.69 &  -7.74 &   37.89\\
 4 & 13:15:42.286 & 42:01:31.89 &  9 &  211 & -11.30 &  -6.69 &  -6.36 &  -7.51 &  -7.38 &  -7.77 &   36.78\\
 5 & 13:15:38.651 & 42:01:42.75 & 12 &  306 & -11.70 &  -7.93 &  -7.58 &  -8.36 &  -8.34 &  -9.21 &   37.82\\
 6 & 13:15:39.155 & 42:01:58.50 & 30 & 1239 & -13.22 &  -6.58 &  -7.25 &  -7.59 &  -7.57 &  -8.06 &   38.13\\
 7 & 13:15:38.886 & 42:02:05.63 & 12 &  507 & -12.25 &  -5.42 &  -5.42 &  -6.68 &  -6.09 &  -7.34 &   38.40\\
 8 & 13:15:39.861 & 42:02:19.51 &  6 &  129 & -10.77 &  -6.52 &  -5.98 &  -6.56 &  -7.27 &  -8.40 &   37.77\\
 9 & 13:15:39.895 & 42:02:19.13 &  9 &  201 & -11.25 &  -7.98 &  -8.24 &  -8.72 &  -8.74 &  -9.07 &   37.58\\
10 & 13:15:40.534 & 42:02:40.14 & 20 &  555 & -12.35 &  -7.11 &  -7.11 &  -7.61 &  -7.81 &  -8.56 &   38.23\\
11 & 13:15:47.401 & 42:02:19.91 & 16 &  175 & -11.09 &  -5.47 &  -5.45 &  -6.74 &  -6.51 &  -7.45 &   37.99\\
12 & 13:15:50.364 & 42:02:25.53 &  6 &  118 & -10.67 &  -7.16 &  -7.68 &  -8.14 &  -8.60 &  -9.43 &   37.59\\
13 & 13:15:53.326 & 42:02:22.53 & 12 &  157 & -10.98 &  -8.44 &  -8.35 &  -9.36 &  -8.49 &  -9.29 & $<  36.47$ \\
14 & 13:15:43.330 & 42:00:57.40 & 25 &  852 & -12.82 &  -6.65 &  -7.08 &  -7.22 &  -7.34 &  -8.49 &   38.73\\
15 & 13:15:56.958 & 42:00:52.14 & 16 &  350 & -11.85 &  -6.54 &  -6.98 &  -8.27 &  -8.20 &  -8.93 &   37.42\\
16 & 13:15:52.281 & 42:01:19.53 & 12 &  346 & -11.84 &  -6.97 &  -7.55 &  -7.86 &  -7.87 &  -7.87 &   38.10\\
NGC 5194 &&&&&&&&&&&\\
 1 & 13:29:54.725 & 47:12:36.45 & 16 &  994 & -12.98 &  -7.76 &  -8.25 &  -8.86 &  -8.67 &  -9.03 &   38.50\\
 2 & 13:29:52.038 & 47:12:42.82 & 20 & 1502 & -13.43 &  -9.31 &  -9.33 &  -9.95 &  -9.46 &  -9.59 &   38.51\\
 3 & 13:29:46.997 & 47:12:23.32 & 12 &  278 & -11.60 &  -6.39 &  -7.25 &  -7.63 &  -7.74 &  -8.84 &   38.03\\
 4 & 13:29:45.231 & 47:11:58.94 &  9 &  118 & -10.67 &  -6.41 &  -6.89 &  -7.69 &  -7.76 &  -8.36 &   37.66\\
 5 & 13:29:44.864 & 47:11:55.18 & 16 &  221 & -11.35 &  -6.49 &  -6.98 &  -7.53 &  -7.80 &  -8.45 &   37.94\\
 6 & 13:29:45.489 & 47:11:55.56 & 12 &   57 &  -9.88 &  -5.88 &  -7.11 &  -7.64 &  -7.68 &  -8.27 &   37.51\\
 7 & 13:29:42.953 & 47:11:05.30 &  9 &   84 & -10.30 &  -6.48 &  -7.20 &  -7.21 &  -7.04 &  -7.84 &   37.82\\
 8 & 13:29:43.359 & 47:10:33.80 &  9 &  132 & -10.79 &  -6.48 &  -6.66 &  -7.02 &  -7.21 &  -8.32 &   37.80\\
 9 & 13:29:54.098 & 47:10:36.07 & 20 &  662 & -12.54 &  -7.14 &  -7.89 &  -8.50 &  -8.49 &  -8.88 &   37.99\\
10 & 13:29:56.526 & 47:10:45.82 & 42 & 2465 & -13.97 &  -8.45 &  -9.23 &  -9.62 &  -9.67 &  -9.81 &   38.58\\
11 & 13:29:56.453 & 47:11:15.82 & 30 &  322 & -11.76 &  -7.26 &  -7.15 &  -7.79 &  -7.70 &  -8.64 &   38.09\\
12 & 13:29:55.828 & 47:11:44.70 & 12 & 1202 & -13.19 &  -8.02 &  -8.47 &  -9.62 &  -9.60 &  -9.42 &   38.56\\
NGC 5248 &&&&&&&&&&&\\
 1 & 13:37:29.691 &  8:54:30.29 & 28 &  466 & -13.71 &  -9.08 &  -9.15 &  -9.71 &  -9.57 & -10.22 &   38.44\\
 2 & 13:37:29.033 &  8:54:20.91 & 24 &  766 & -14.25 &  -8.95 &  -9.59 & -10.10 & -10.25 & -10.46 &   38.53\\
 3 & 13:37:28.780 &  8:53:27.66 & 42 & 1120 & -14.66 & -10.00 & -10.68 & -10.86 & -10.79 & -10.98 &   38.93\\
 4 & 13:37:29.286 &  8:53:14.91 & 15 &  278 & -13.15 &  -8.29 &  -8.21 &  -8.60 &  -8.94 &  -9.86 &   38.60\\
 5 & 13:37:30.374 &  8:52:50.16 & 20 &  140 & -12.40 &  -8.65 &  -8.78 &  -9.33 &  -8.91 &  -9.65 &   38.37\\
 6 & 13:37:32.601 &  8:52:43.41 & 20 &  157 & -12.53 &  -8.85 &  -8.81 &  -9.35 &  -9.51 & -10.39 &   38.23\\
 7 & 13:37:32.930 &  8:52:32.54 & 24 &  291 & -13.20 &  -9.41 &  -9.65 & -10.42 & -10.10 & -10.46 &   39.02\\
 8 & 13:37:34.929 &  8:53:00.29 & 15 &  263 & -13.09 &  -8.94 &  -9.19 &  -9.55 &  -9.51 &  -9.94 &   38.65\\
 9 & 13:37:34.575 &  8:53:06.29 & 16 &  208 & -12.83 &  -8.54 &  -8.90 &  -8.64 &  -9.02 &  -9.70 &   38.46\\
10 & 13:37:35.106 &  8:52:20.54 & 12 &  157 & -12.53 &  -8.55 &  -7.88 &  -8.69 &  -8.71 &  -9.34 &   38.50\\
11 & 13:37:33.967 &  8:53:12.29 &  6 &   40 & -11.04 &  -7.56 &  -7.71 &  -7.95 &  -8.58 &  -9.27 &   37.61\\
12 & 13:37:31.386 &  8:53:37.04 & 42 &  315 & -13.29 &  -9.24 &  -9.26 &  -9.55 &  -9.45 & -11.57 &   38.75\\
13 & 13:37:31.285 &  8:53:31.41 & 30 &  155 & -12.52 &  -9.15 &  -9.28 &  -9.72 &  -9.77 & -10.32 &   38.55\\
NGC 5457 &&&&&&&&&&&\\
 1 & 14:03:07.750 & 54:22:43.86 & 20 &  227 & -10.46 &  -5.71 &  -5.56 &  -6.35 &  -7.35 &  -7.76 &   37.62\\
 2 & 14:03:05.818 & 54:22:45.35 & 12 &  108 &  -9.65 &  -5.75 &  -5.14 &  -5.28 &  -5.93 &  -6.90 & $<  37.03$ \\
 3 & 14:03:04.745 & 54:22:45.73 & 16 &  150 & -10.01 &  -5.73 &  -5.22 &  -5.93 &  -6.09 &  -7.50 & $<  37.66$ \\
 4 & 14:03:03.843 & 54:22:47.22 & 25 &  246 & -10.55 &  -6.51 &  -6.58 &  -7.17 &  -7.12 &  -7.76 & $<  37.76$ \\
 5 & 14:03:03.285 & 54:22:49.09 &  6 &   52 &  -8.85 &  -5.29 &  -4.76 &  -6.12 &  -6.33 &  -6.43 & $<  37.45$ \\
 6 & 14:03:03.198 & 54:23:01.84 & 20 &  206 & -10.35 &  -6.13 &  -6.24 &  -7.28 &  -7.92 &  -8.44 &   36.22\\
 7 & 14:03:29.293 & 54:22:02.54 & 25 &  270 & -10.65 &  -6.86 &  -6.85 &  -7.23 &  -7.46 &  -7.95 &   37.41\\
 8 & 14:03:28.604 & 54:21:44.92 & 81 & 1251 & -12.31 &  -7.04 &  -6.94 &  -7.69 &  -7.38 &  -8.16 &   38.12\\
 9 & 14:03:27.230 & 54:21:27.31 & 25 &  217 & -10.41 &  -6.86 &  -6.97 &  -7.51 &  -7.47 &  -7.89 &   38.18\\
10 & 14:03:26.243 & 54:21:25.07 & 35 &  386 & -11.04 &  -7.27 &  -7.50 &  -7.84 &  -7.99 &  -8.37 &   37.83\\
11 & 14:03:20.620 & 54:20:47.22 & 24 &  204 & -10.35 &  -6.23 &  -6.61 &  -7.02 &  -7.00 &  -7.54 &   37.81\\
12 & 14:03:19.032 & 54:20:18.73 & 12 &   98 &  -9.55 &  -6.37 &  -6.55 &  -7.00 &  -7.19 &  -7.47 &   37.35\\
13 & 14:03:12.514 & 54:19:07.87 & 36 &  672 & -11.64 &  -6.43 &  -6.19 &  -7.12 &  -6.76 &  -7.79 &   37.94\\
14 & 14:03:14.186 & 54:19:08.99 & 15 &  150 & -10.01 &  -6.34 &  -6.16 &  -6.95 &  -6.72 &  -6.89 &   38.49\\
15 & 14:02:57.854 & 54:19:20.93 & 12 &   53 &  -8.88 &  -5.93 &  -5.53 &  -5.78 &  -6.05 &  -6.85 &   37.52\\
16 & 14:02:58.199 & 54:19:07.06 &  9 &   65 &  -9.10 &  -5.90 &  -5.18 &  -6.56 &  -6.57 &  -6.93 &   37.64\\
17 & 14:03:27.177 & 54:19:46.81 & 56 & 1118 & -12.19 &  -7.32 &  -7.53 &  -8.13 &  -8.02 &  -8.45 &   38.24
\enddata
\tablenotetext{a}{Upper limits on the H$\alpha$ flux result when the HII region is not seen at
the position of the $3.6\mu$m source.}
\label{results}
\end{deluxetable}

\clearpage

\begin{deluxetable}{lcccccc}
\tabletypesize{\scriptsize}  \tablecaption{Properties of Embedded Sources}
\tablehead{
\colhead{Galaxy} &
\colhead{Region} &
\colhead{$\log_{10}M_{H\alpha}$\tablenotemark{a}}&
\colhead{$A_{\rm g}(H\alpha)$\tablenotemark{a}}&
\colhead{$\log_{10}M_{3.6}$} &
\colhead{$A_{\rm g}(3.6\mu{\rm m})$} &
\colhead{$\Delta A_{\rm g}$\tablenotemark{a}}\\
&&
\colhead{($M_\odot$)} &
\colhead{(mag)} &
\colhead{($M_\odot$)} &
\colhead{(mag)} &
\colhead{(mag)}
}
\startdata
NGC 4321 &  1 &  3.8 & $ 2.5\pm  0.4$ &  5.3 & $ 6.5\pm  0.4$ &  4.2\\
 &  2 &  4.0 & $ 2.7\pm  0.4$ &  5.0 & $ 5.3\pm  0.4$ &  2.8\\
 &  3 &  3.5 & $ 1.9\pm  0.5$ &  5.3 & $ 6.4\pm  0.5$ &  4.8\\
 &  4 & $\sim 3.0$ & $> 0.2\pm  0.5$ &  5.0 & $       5.3\pm  0.5$ & $ \sim       5.3$\\
 &  5 &  3.4 & $ 0.9\pm  0.3$ &  5.3 & $ 5.7\pm  0.3$ &  5.1\\
 &  6 &  3.4 & $ 1.9\pm  0.5$ &  5.3 & $ 6.5\pm  0.5$ &  4.8\\
 &  7 &  3.6 & $ 2.0\pm  0.4$ &  5.3 & $ 6.1\pm  0.4$ &  4.4\\
 &  8 & $\sim 2.0$ & $>-1.3\pm  0.3$ &  4.4 & $       4.7\pm  0.3$ & $ \sim       6.2$\\
 &  9 & $\sim 2.2$ & $>-1.1\pm  0.8$ &  4.4 & $       4.3\pm  0.8$ & $ \sim       5.7$\\
 & 10 & $\sim 2.7$ & $>-0.2\pm  0.8$ &  5.0 & $       5.5\pm  0.8$ & $ \sim       6.0$\\
 & 11 &  4.0 & $ 3.1\pm  0.5$ &  6.2 & $ 8.5\pm  0.5$ &  5.7\\
 & 12 &  3.9 & $ 2.1\pm  0.3$ &  5.8 & $ 6.9\pm  0.3$ &  5.1\\
 & 13 &  3.7 & $ 2.4\pm  0.5$ &  5.5 & $ 6.9\pm  0.5$ &  4.7\\
 & 14 &  3.0 & $ 0.3\pm  0.4$ &  4.5 & $ 4.0\pm  0.4$ &  4.0\\
 & 15 &  3.7 & $ 1.0\pm  0.9$ &  5.6 & $ 5.8\pm  0.9$ &  5.0\\
NGC 5055 &  1 & $\sim 2.4$ & $>-1.9\pm   0.8$ &  3.7 & $ 1.5\pm  0.8$ & $\sim 3.3$\\
 &  2 &  2.5 & $-2.9\pm  0.6$ &  4.5 & $ 1.9\pm  0.6$ &  5.1\\
 &  3 &  3.2 & $ 1.4\pm  0.7$ &  5.0 & $ 6.0\pm  0.7$ &  4.8\\
 &  4 &  2.1 & $-2.2\pm  0.6$ &  4.5 & $ 3.8\pm  0.6$ &  6.4\\
 &  5 &  3.1 & $-0.8\pm  0.5$ &  4.6 & $ 3.0\pm  0.5$ &  4.0\\
 &  6 &  3.4 & $ 0.9\pm  0.5$ &  5.2 & $ 5.4\pm  0.5$ &  4.9\\
 &  7 &  3.7 & $ 2.8\pm  0.6$ &  4.8 & $ 5.6\pm  0.6$ &  3.1\\
 &  8 &  3.1 & $ 0.4\pm  0.7$ &  4.3 & $ 3.4\pm  0.7$ &  3.3\\
 &  9 &  2.9 & $-1.6\pm  0.4$ &  4.5 & $ 2.4\pm  0.4$ &  4.3\\
 & 10 &  3.5 & $ 0.9\pm  0.4$ &  4.8 & $ 4.1\pm  0.4$ &  3.5\\
 & 11 &  3.3 & $ 1.6\pm  0.7$ &  4.3 & $ 4.1\pm  0.7$ &  2.8\\
 & 12 &  2.9 & $-1.3\pm  0.7$ &  4.1 & $ 1.9\pm  0.7$ &  3.4\\
 & 13 & $\sim 1.8$ & $>-4.7\pm  0.4$ &  4.3 & $       1.6\pm  0.4$ & $ \sim       6.5$\\
 & 14 &  4.0 & $ 2.4\pm  0.4$ &  5.1 & $ 5.1\pm  0.4$ &  3.0\\
 & 15 &  2.7 & $-1.3\pm  0.8$ &  4.7 & $ 3.8\pm  0.8$ &  5.3\\
 & 16 &  3.4 & $ 0.6\pm  0.5$ &  4.7 & $ 3.9\pm  0.5$ &  3.6\\
NGC 5194 &  1 &  3.8 & $ 0.7\pm  0.5$ &  5.2 & $ 4.1\pm  0.5$ &  3.4\\
 &  2 &  3.8 & $-0.3\pm  0.4$ &  5.3 & $ 3.6\pm  0.4$ &  4.1\\
 &  3 &  3.3 & $ 0.4\pm  0.7$ &  4.6 & $ 3.7\pm  0.7$ &  3.5\\
 &  4 &  2.9 & $-0.3\pm  0.6$ &  4.2 & $ 2.8\pm  0.6$ &  3.4\\
 &  5 &  3.2 & $ 0.4\pm  0.6$ &  4.4 & $ 3.4\pm  0.6$ &  3.3\\
 &  6 &  2.8 & $-0.6\pm  0.8$ &  3.9 & $ 2.1\pm  0.8$ &  3.0\\
 &  7 &  3.1 & $ 0.4\pm  0.3$ &  4.1 & $ 2.9\pm  0.3$ &  2.8\\
 &  8 &  3.1 & $ 0.3\pm  0.5$ &  4.3 & $ 3.3\pm  0.5$ &  3.2\\
 &  9 &  3.3 & $-0.2\pm  0.6$ &  5.0 & $ 4.0\pm  0.6$ &  4.5\\
 & 10 &  3.9 & $ 0.1\pm  0.5$ &  5.5 & $ 4.1\pm  0.5$ &  4.4\\
 & 11 &  3.4 & $ 0.5\pm  0.4$ &  4.7 & $ 3.8\pm  0.4$ &  3.6\\
 & 12 &  3.8 & $ 0.3\pm  0.8$ &  5.3 & $ 3.9\pm  0.8$ &  3.8\\
NGC 5248 &  1 &  3.7 & $-0.5\pm  0.4$ &  5.4 & $ 3.7\pm  0.4$ &  4.2\\
 &  2 &  3.8 & $-0.6\pm  0.6$ &  5.6 & $ 3.9\pm  0.6$ &  4.7\\
 &  3 &  4.2 & $-0.4\pm  0.4$ &  5.8 & $ 3.6\pm  0.4$ &  4.2\\
 &  4 &  3.9 & $ 0.7\pm  0.5$ &  5.2 & $ 4.1\pm  0.5$ &  3.7\\
 &  5 &  3.7 & $-0.2\pm  0.3$ &  4.9 & $ 2.9\pm  0.3$ &  3.4\\
 &  6 &  3.5 & $-0.8\pm  0.5$ &  4.8 & $ 2.4\pm  0.5$ &  3.5\\
 &  7 &  4.3 & $ 0.5\pm  0.5$ &  5.2 & $ 2.7\pm  0.5$ &  2.5\\
 &  8 &  3.9 & $ 0.1\pm  0.3$ &  5.2 & $ 3.4\pm  0.3$ &  3.5\\
 &  9 &  3.7 & $ 0.1\pm  0.3$ &  5.0 & $ 3.3\pm  0.3$ &  3.5\\
 & 10 &  3.8 & $ 0.6\pm  0.4$ &  5.0 & $ 3.6\pm  0.4$ &  3.3\\
 & 11 &  2.9 & $-1.2\pm  0.5$ &  4.3 & $ 2.1\pm  0.5$ &  3.6\\
 & 12 &  4.0 & $ 0.0\pm  0.6$ &  5.2 & $ 2.9\pm  0.6$ &  3.1\\
 & 13 &  3.8 & $-0.3\pm  0.4$ &  4.3 & $ 1.0\pm  0.4$ &  1.6\\
NGC 5457 &  1 &  2.9 & $ 0.4\pm  0.8$ &  4.0 & $ 3.3\pm  0.8$ &  2.9\\
 &  2 & $\sim 2.3$ & $>-0.3\pm  0.5$ &  3.7 & $       3.3\pm  0.5$ & $ \sim       3.8$\\
 &  3 & $\sim 2.9$ & $> 1.0\pm  0.6$ &  4.0 & $       3.6\pm  0.6$ & $ \sim       2.8$\\
 &  4 & $\sim 3.0$ & $> 0.3\pm  0.4$ &  4.1 & $       3.1\pm  0.4$ & $ \sim       3.0$\\
 &  5 & $\sim 2.7$ & $> 0.8\pm  0.7$ &  3.5 & $       2.7\pm  0.7$ & $ \sim       2.2$\\
 &  6 &  1.5 & $-3.7\pm  0.9$ &  3.8 & $ 2.1\pm  0.9$ &  6.1\\
 &  7 &  2.7 & $-0.8\pm  0.4$ &  4.1 & $ 2.8\pm  0.4$ &  3.9\\
 &  8 &  3.4 & $ 0.8\pm  0.4$ &  4.9 & $ 4.5\pm  0.4$ &  4.0\\
 &  9 &  3.5 & $ 1.0\pm  0.4$ &  4.0 & $ 2.5\pm  0.4$ &  1.7\\
 & 10 &  3.1 & $-0.3\pm  0.4$ &  4.2 & $ 2.6\pm  0.4$ &  3.1\\
 & 11 &  3.1 & $ 0.6\pm  0.4$ &  4.0 & $ 2.9\pm  0.4$ &  2.6\\
 & 12 &  2.6 & $-0.6\pm  0.4$ &  3.7 & $ 2.1\pm  0.4$ &  3.0\\
 & 13 &  3.2 & $ 0.9\pm  0.5$ &  4.6 & $ 4.4\pm  0.5$ &  3.8\\
 & 14 &  3.8 & $ 2.5\pm  0.4$ &  4.0 & $ 3.1\pm  0.4$ &  0.9\\
 & 15 &  2.8 & $ 0.7\pm  0.3$ &  3.4 & $ 2.2\pm  0.3$ &  1.7\\
 & 16 &  2.9 & $ 0.8\pm  0.6$ &  3.4 & $ 2.1\pm  0.6$ &  1.6\\
 & 17 &  3.5 & $ 0.7\pm  0.4$ &  4.8 & $ 3.9\pm  0.4$ &  3.5
\enddata
\tablenotetext{a}{Approximation sign is given for $3.6\mu$m sources with upper limits to the
H$\alpha$ flux. In these cases, the extinction at $g$-band is a lower limit because optical
emission in the SDSS bands is not present. The H$\alpha$ could also be a lower limit
if the extinction is much higher than its tabulated limit, or it could be an upper limit
if, for example, the source is old and there is no H$\alpha$ at all. Finally, in these same
cases, the extinction difference between H$\alpha$ and $3.6\mu$m shown in column 7
would be an upper limit if the H$\alpha$ extinction is a lower limit. In fact, the similarity of the
extinction difference in column 7 and the extinction for the $3.6\mu$m emission alone in column 6, combined with the
near-zero extinction difference between the H$\alpha$ and the SDSS bands (column 3) implies that the
broad-band optical emission is close to its upper limit and the H$\alpha$ and broadband light are
extincted by similar amounts.
}

\label{results2}
\end{deluxetable}

\clearpage

\begin{deluxetable}{lccccc}
\tabletypesize{\scriptsize}  \tablecaption{Average Complex Masses and Extinctions}
\tablehead{
\colhead{Galaxy} &
\colhead{$<\log_{10}$ Mass$\,>$ $(M_\odot)$} &
\colhead{$<A_{\rm g}>$ (mag)} &
\colhead{$<\log_{10}$ Mass$\,>$ $(M_\odot)$} &
\colhead{$<A_{\rm g}>$ (mag)} &
\colhead{$\Delta A_{\rm g}$}
}
\startdata
NGC 4321  & $  3.6\pm  0.3$ & $  1.9\pm  0.8$ & $  5.2\pm  0.5$ & $  5.9\pm  1.1$ & $  4.6\pm  0.7$\\
NGC 5055  & $  3.1\pm  0.5$ & $  0.1\pm  1.7$ & $  4.6\pm  0.4$ & $  3.6\pm  1.4$ & $  4.1\pm  1.0$\\
NGC 5194  & $  3.4\pm  0.4$ & $  0.1\pm  0.4$ & $  4.7\pm  0.5$ & $  3.5\pm  0.6$ & $  3.6\pm  0.5$\\
NGC 5248  & $  3.8\pm  0.3$ & $ -0.2\pm  0.5$ & $  5.1\pm  0.4$ & $  3.0\pm  0.8$ & $  3.5\pm  0.8$\\
NGC 5457  & $  3.0\pm  0.5$ & $  0.2\pm  1.4$ & $  4.0\pm  0.4$ & $  3.0\pm  0.7$ & $  3.0\pm  1.3$
\enddata
\label{averages}
\end{deluxetable}

\clearpage
\begin{figure}
\centering
\includegraphics[width=5.5in]{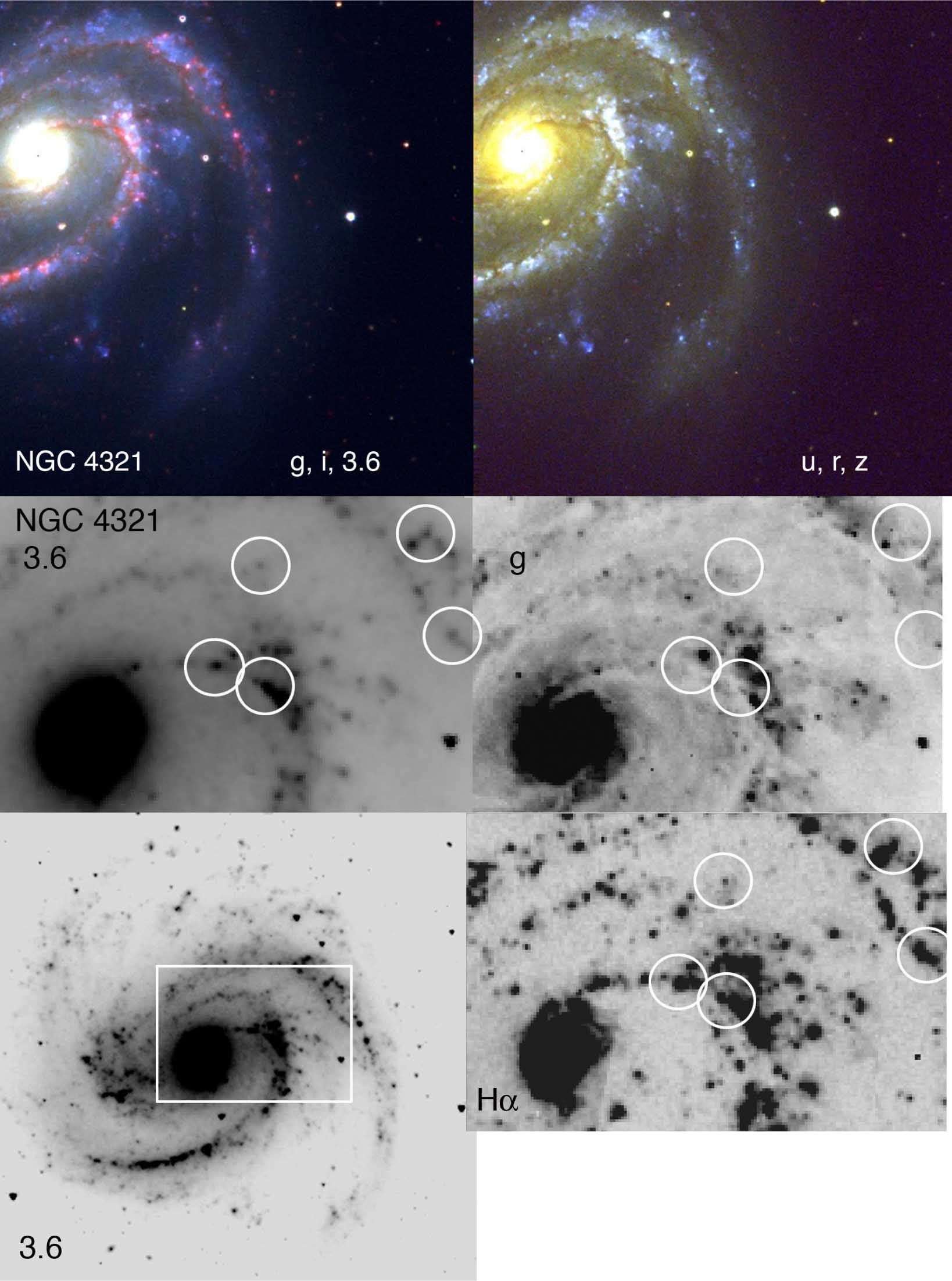}
\caption{(Top left) color composite of NGC 4321 with SDSS $g$-band image for blue,
SDSS $i$-band image for green, and IRAC $3.6\,\mu$m image for red, all with
$0.75^{\prime\prime}$ pixels and combined by rotating, stretching, and aligning the
optical images to match the infrared images. (Top right) SDSS color image
remade from the {\it u}, {\it r}, and {\it z} SDSS filters
using data from the SDSS webpage. (Middle left) sample region at $3.6\,\mu$m
with circles highlighting $3.6\,\mu$m sources that do
not appear in optical bands. (Middle right) The same circles
on an SDSS $g$-band image (the circles are $\sim10\times$
larger than the measurement rectangles). (Bottom left) $3.6\mu$m
full image with a rectangle indicating the sample region. (Bottom right)
Sample region again in H$\alpha$.
}\label{N4321montagenew2}\end{figure}


\clearpage
\begin{figure}
\centering
\includegraphics[width=6.5in]{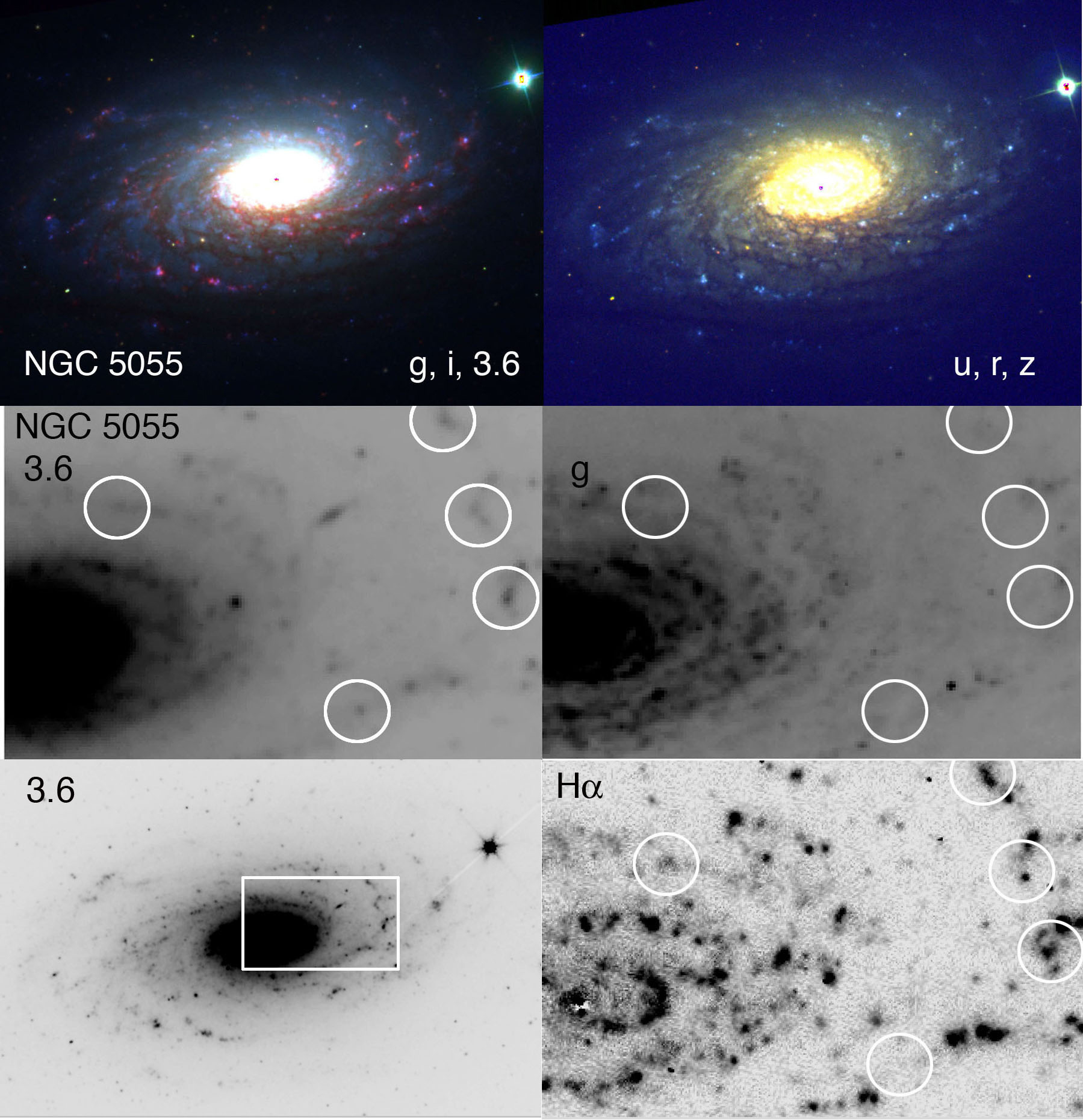}
\caption{NGC 5055, as in Figure 1.}\label{N5055colormontagenew}\end{figure}


\clearpage
\begin{figure}
\centering
\includegraphics[width=6.5in]{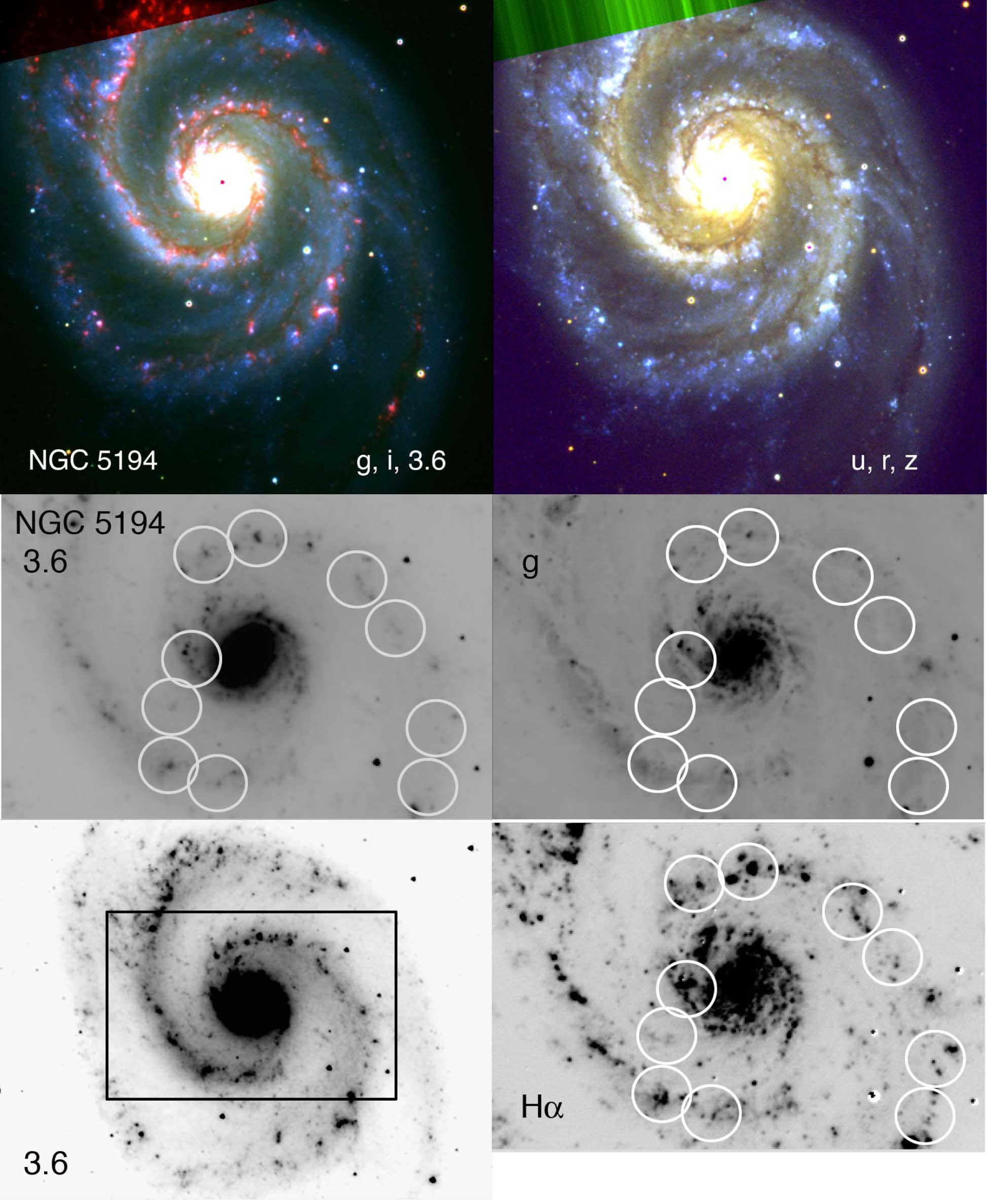}
\caption{NGC 5194, as in Figure 1.}\label{N5194montagenew}\end{figure}


\clearpage
\begin{figure}
\centering
\includegraphics[width=6.5in]{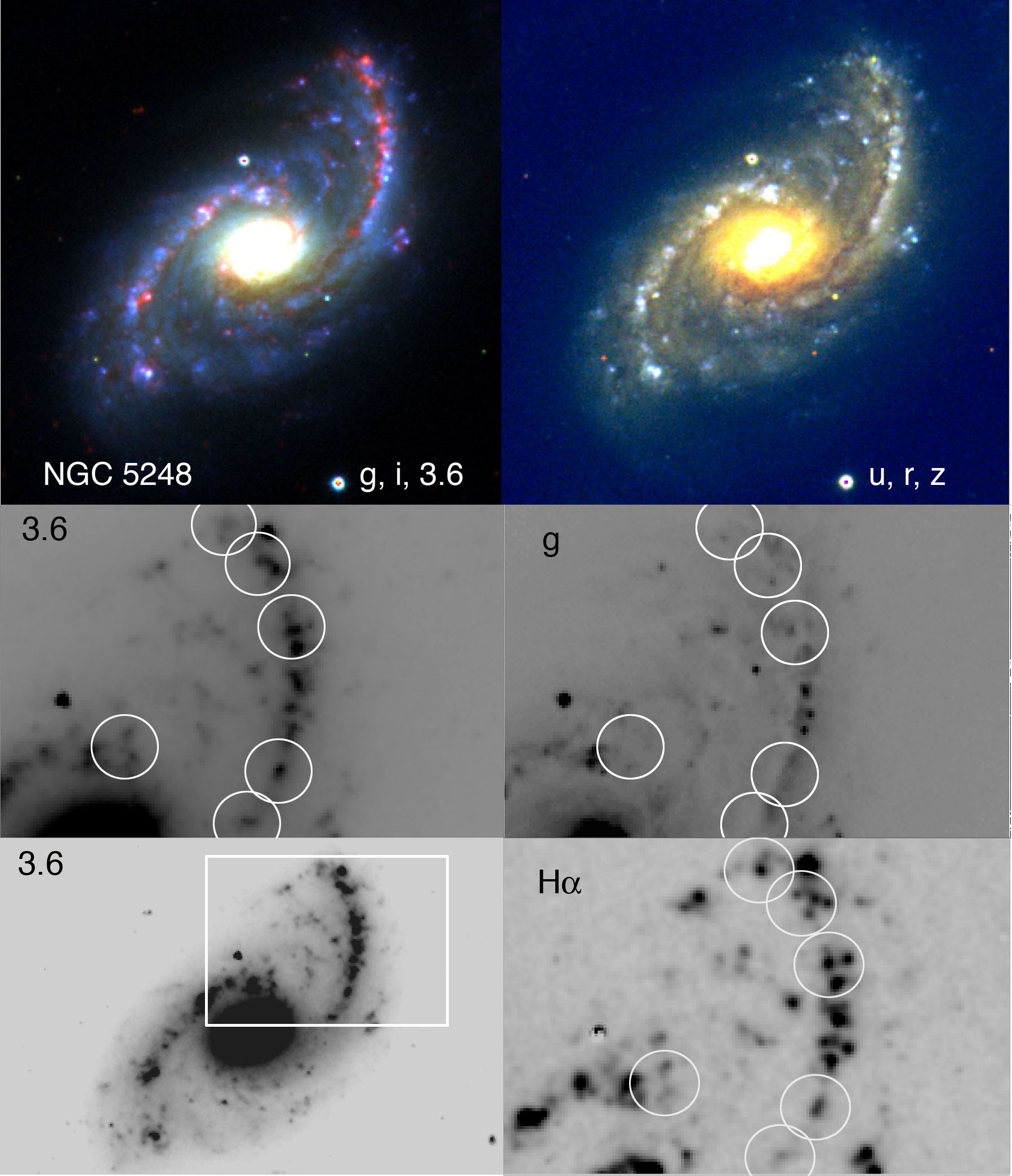}
\caption{NGC 5248, as in Figure 1.}\label{N5248color_SDSS_ch1_g-121712}\end{figure}

\clearpage
\begin{figure}
\centering
\includegraphics[width=6.5in]{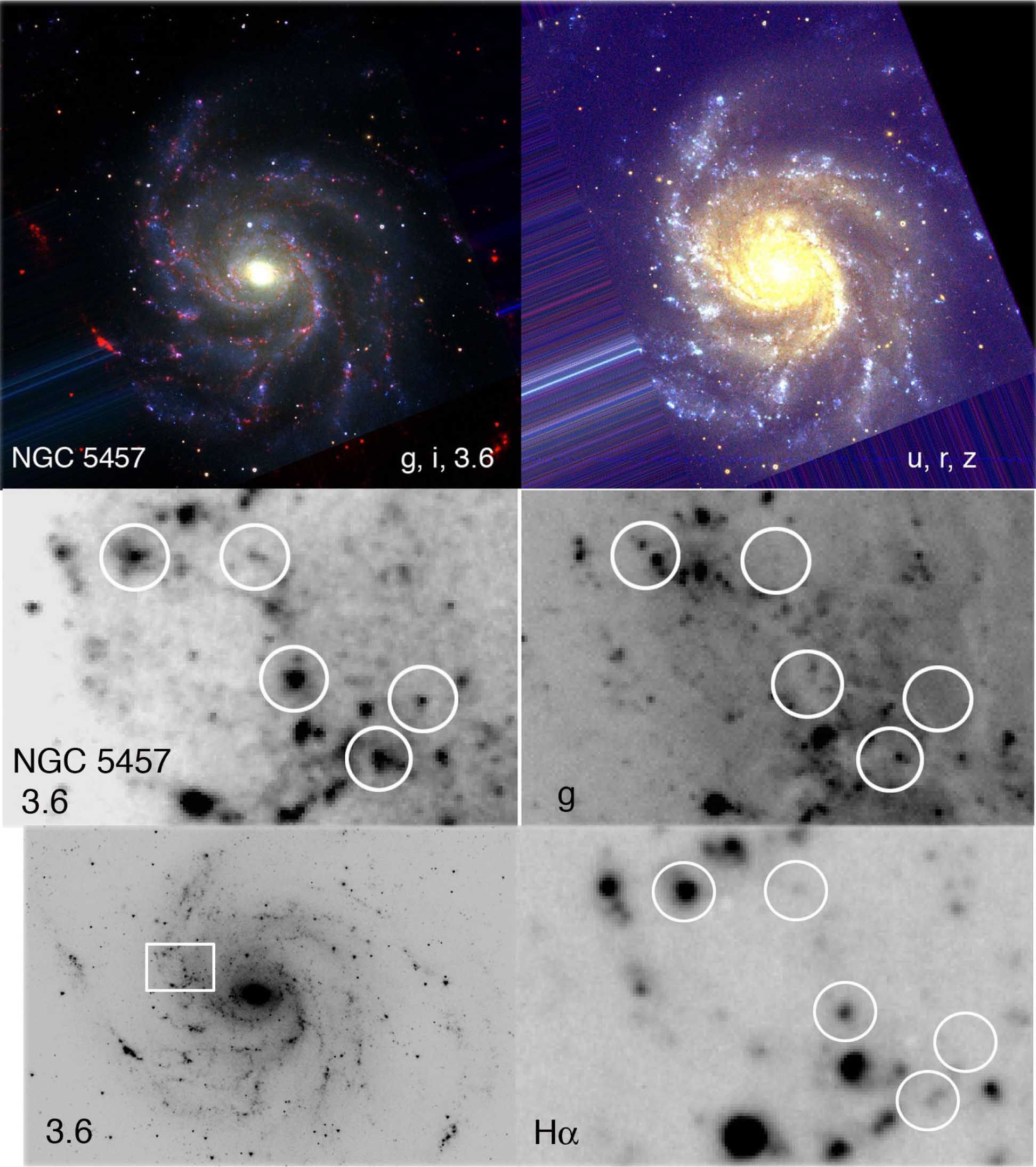}
\caption{NGC 5457, as in Figure 1.}\label{N5457montagenew}\end{figure}


\end{document}